\documentclass[11pt,twoside]{article}

\usepackage{asp2006}
\usepackage{graphicx}
\usepackage{enumerate}
\usepackage{lscape}
\begin{document}
\title{X-ray echoes of infrared flaring in Sgr A*}
\author{Mark Wardle}
\affil{Department of Physics \& Engineering, Macquarie University, NSW 
2109, Australia}

\begin{abstract}
  
Sgr A* exhibits flaring in the infrared several times each day,
occasionally accompanied by flaring in X-rays.  The infrared flares
are believed to arise through synchrotron emission from a transient
population of accelerated electrons.  The X-ray flaring has been
interpreted as self-synchrotron-compton, inverse compton, or
synchrotron emission associated with the transient electrons.

Here I consider the upscattering of infrared flare photons by
relativistic thermal electrons in the accretion flow around Sgr A*.
Typical profiles of electron density and temperature in the accretion
flow are adopted and the X-ray light curves produced by upscattering
of infrared flare photons by the accretion flow are computed.

Peak X-ray luminosities between $10^{33}$ and $10^{34}$\,erg\,s$^{-1}$
are attained for a 10\,mJy near-infrared flare, compatible with
observed coincident infrared/X-ray flares from Sgr A*.  Even if this process
is not responsible for the observed flares it still presents a serious
constraint on accretion flow models, which must avoid over-producing
X-rays and also predicting observable time lags between flaring in infrared
and in X-rays.

Future high-resolution infrared instrumentation will be able to place the
location of the infrared flare and in coordination with the X-ray
would severely constrain the disc geometry and the radial profiles of
electron density and temperature in the accretion flow.

\end{abstract}

\section{Introduction}

The quasi-steady emission from Sgr A* spanning radio to submillimeter
frequencies is believed to be synchrotron radiation from electrons
participating in an accretion flow that extends out to radii of
100\,AU or more from the event horizon (e.g.\ Yuan et al.\ 2003), or
alternatively close the base of a jet (e.g.\ Falcke \& Markoff 2000).  The
electron temperature and magnetic field strength in the inner regions
of the flow are inferred to be $kT_e\sim 10$\,MeV and $\sim 30$\,G,
respectively.

Several times a day, Sgr A* exhibits dramatic flaring in the
near-infrared, with $\sim30$-minute flare durations and peak fluxes
(corrected for extinction) ranging downwards from 30\,mJy (e.g.\ Genzel
et al.\ 2003; Eckart et al.\ 2006b; Yusef-Zadeh et al.\ 2006, 2009; Do
et al.\ 2009).  The emission is highly polarized (Eckart et al.\
2006a; Meyer et al.\ 2006; Trippe et al.\ 2007) and so the flaring is
almost certainly synchrotron emission from a transient population of
$\sim$GeV electrons that are promptly accelerated by a violent event
and subsequently cool by synchrotron losses.

A small fraction of infrared flares are accompanied by an X-ray flare
which peaks within a few minutes of (and possibly contemporaneously
with) the near-infrared (e.g.\ Baganoff et al.\ 2001; Porquet et al.\
2008).  Several scenarios for the X-ray flaring have been considered: 
\begin{enumerate}[(i)]
\setlength{\itemsep}{-2pt}
\item upscattering of sub-millimeter synchrotron photons by the
transient population of GeV electrons that emitted them, ie.\
self-synchrotron-compton (e.g.\ Yuan, Quataert \& Narayan 2004; Eckart et 
al.\ 2004); 
\item upscattering of sub-millimeter photons from
the accretion flow by the transient GeV electrons (Yusef-Zadeh et 
al.\ 2006); 
\item upscattering of infrared flare photons
by the $\sim$10 MeV electrons in the accretion flow (Yusef-Zadeh et 
al.\ 2006, 2009); and
\item synchrotron emission from the high energy tail of the accelerated electron
population (e.g.\ Markoff et al.\ 2001; Yuan, Quataert \& Narayan 
2003, 2004).
\end{enumerate}

Dodds-Eden et al.\ (2009) argue that the first three scenarios are
unable to produce the observed X-ray fluxes for reasonable parameters.
The synchrotron-self-compton scenario requires photons to be produced
and subsequently upscattered all within the same infrared source
region.  The requisite optical depth to Thomson scattering implies a
source region that is unreasonably compact and overpressured relative
to its surroundings, with equipartition magnetic field strengths of at
least 2\,kG. Similarly, in the second scenario, the predicted X-ray
flux falls short unless the upscattering rate is boosted by adopting
an unreasonably compact millimeter/infrared source region, with an
equipartition field strength in excess of 200\,G. The third scenario
is excluded on the grounds that it should produce a similar X-ray flux
as the second.  This follows from a fundamental symmetry between
scenarios (ii) and (iii): if one chooses any electron from the infrared
source region and any from the millimetre-emitting population, then
synchrotron photons emitted by one and upscattered by the other will
have the same energy; and the cross-section for scattering in both
cases is identical (i.e.\ the Thomson cross section).  So then two
distinct populations of electrons will produce similar X-ray fluxes by
upscattering the synchrotron emission from the other.  In addition,
Dodds-Eden et al.\ (2009) note that this scenario implies time-delays
between infrared and X-rays and these are strongly constrained by
existing data.  Instead, they favour synchrotron emission from a
high-energy tail of the transient electrons producing the infrared
emission.

So are inverse compton scenarios ruled out?  No, at least not scenario
(iii)!  As noted in Yusef-Zadeh et al.\ (2009), the symmetry argument
outlined above implicitly assumes that both electron populations are
optically thin to seed and upscattered photons.  In contrast, the
accretion flow is optically thick to millimeter photons and optically
thin to infrared and X-ray photons.  Thus many of the millimeter
synchrotron photons are reabsorbed and the upscattered flux in (ii) is
reduced by a factor of $\tau$ relative to scenario (iii).  Indeed,
Yusef-Zadeh et al.\ (2009) estimated that the X-ray flux produced by
(iii) should be comparable to that observed.

Regardless of the mechanism actually responsible for X-ray flares from
Sgr A*, upscattering of infrared flare photons by the accretion flow
is efficient enough to constrain the electron density and temperature
profiles in accretion models, which must avoid over-producing X-rays
and also avoid contradicting the limits on the time lag between
flaring in infrared and in X-rays.

To illustrate this, I consider the X-ray emission produced by
inverse-compton scattering of infrared flare photons from simple (but
reasonable) accretion flow profiles.

\section{Calculations}

Some justifiable approximations simplify the calculation of the
X-ray echo.  First, the energy of an upscattered photon with initial
energy $h\nu_\mathrm{infrared}$ is assumed to be $\gamma^2 h\nu_\mathrm{infrared}$
where $\gamma$ is the electron Lorentz factor; the total production
rate of upscattered photons per unit volume is $n_\mathrm{infrared} n_e
\sigma_T c$ where $n_\mathrm{infrared}$ and $n_e$ are the number densities
of infrared photons and relativistic electrons; and the upscattering is
assumed to be isotropic.  Second, the electron population is
characterised by an approximate relativistic maxwellian $f(x) = 1/2 x^2
\exp(-x^2)$ where $x = E / kT_e$ (valid for $kT_e \ga 2$\,MeV).

\begin{figure}
    \centering
     \includegraphics[scale=0.5]{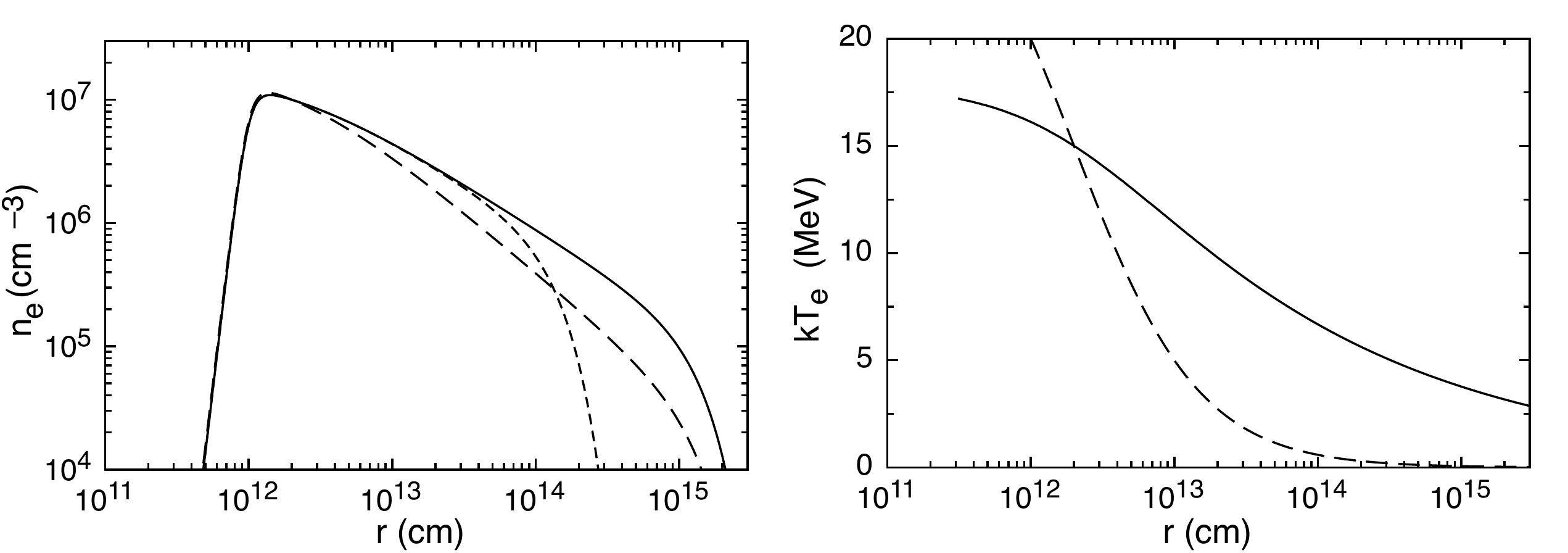}
    \caption{\textit{(Left)} Adopted midplane electron density 
    profiles, with $n_e\propto r^{-0.75}$ (solid and short dashed 
    curves) or $r^{-1}$ with cutoffs at 500 $R_s$ and 50 $R_s$ (solid or dashed 
    curve respectively). \textit{(Right)} Adopted radial profiles of 
    electron temperature, with $T_e \propto r^{-0.25}$ or $T_e 
    \propto r^{-1}$ (solid or dashed curves respectively). 
    \label{fig:nTprofile}
    }
\end{figure}

The adopted model profiles for the electron density and temperature
are $n_e \propto r^{-0.75}$, $T_e \propto r^{-0.25}$ or $n_e \propto
r^{-1}$, $T_e \propto r^{-1}$.  The density is truncated within
$2\,R_s$ and beyond $50\,R_s$ and is assumed to be either spherically
symmetric or disk-like with $h/r = 0.5$.  The density and temperature
profiles are plotted in Fig.\ \ref{fig:nTprofile}.  These numbers are
within the typical ranges considered in analytic estimates (e.g.\ Loeb
\& Waxman 2007, semi-analytic models for the accretion flow (e.g.
Yuan, Quataert \& Narayan 2003); and MHD simulations (e.g\
Mo\'scibrodzka et al.\ 2009).

For the sake of illustration, we consider an infrared flare with a
peak flux of 10\,mJy at $2\micron$, a gaussian light curve with a 30
minute FWHM, and a $\lambda^{0.5}$ spectrum extending up to
$0.2\micron$.  The infrared emission is assumed to originate from a point
source at a specified location.  Note that the X-ray echo depends linearly on
the flare photons, and an extended flare emission region could easily
be modelled as a sum of the responses from a cluster of point sources.
Here we focus on this simple model flare and consider a few different
positions of the infrared source and explore the effect of changing some of
the parameters of the accretion flow.

\section{Results}
Fig.\ 2 shows the X-ray response for a variety of infrared flare locations
and choices of accretion-flow
profiles.  
\begin{figure}
    \centering
     \includegraphics[scale=0.55]{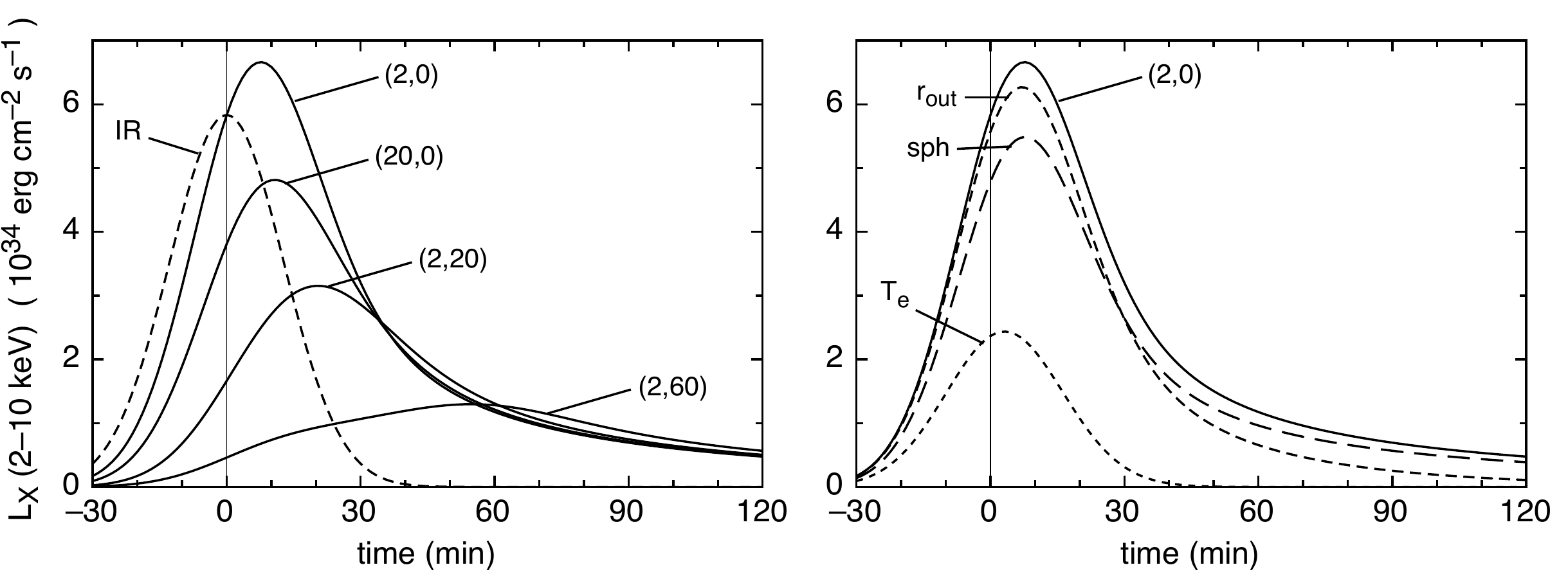}
    \caption{X-ray light curves resulting from inverse Compton
    scattering of infrared photons produced by a compact infrared flare
    occuring in the vicinity of Sgr A*.  Left panel -- solid curves show the
    resultant X-ray light curves for different locations of the infrared
    source region, labelled by $(r,z)$ (in units of $10^{12}$\,cm).
    The dashed curve shows the assumed gaussian profile of the infrared
    flare which has a 10\,mJy peak at t=0 and FWHM 30 minutes.  Right 
    panel
    -- broken curves show additional light curves for $(r,z)=(2,0)$
    with differing assumptions about the accretion flow: $r_out$
    reduced from 500$R_s$ to 50$R_s$ (\emph{short dashes});
    spherically symmetric with the same total number of electrons
    (\emph{long dashes}); $T_e \propto r^{-1}$ (\emph{dotted}).
    \label{fig:xrays}
    }
\end{figure}
First we adopt the density and temperature profile indicated by the
solid curves in the left and right panels of Fig.\ 1 respectively and
a disc scale height $h/r = 0.5$.  The left panel of Fig.\ 2 shows the
resulting X-ray flare for different choices of infrared source positions.
The central concentration of the electron density near the inner edge
of the accretion flow and the centrally-concentrated temperature
profile means that there is a delay in the peak response of the
upscattered X-rays.  As the source location is moved successively
further away from the inner edge of the disk the delay and width of
the profile both increase.  The peak flux is correspondingly
diminished.

The right hand panel of Fig.\ 2 shows the effect of varying the
adopted accretion flow while keeping the infrared source region at a fixed
location.  The upscattered X-ray light curve is relatively insensitive
to changing the accretion flow outer boundary and aspect ratio
(short-dashed and long-dashed curves respectively), but is sensitive
to changes to the temperature profile (dotted curve) as the electron
temperature controls the fraction of the electrons that are able to
upscatter the infrared photons to X-ray energies.  In particular, switching
from the $T_e\propto r^{-0.25}$ to a steeper $T_e\propto r^{-1}$
profile (solid vs dashed curves in Fig.\ 1) means that less electrons
overall have sufficient energy to upscatter infrared photons to X-ray
energies, so the peak X-ray flux and total fluence are both reduced.
The energetic electrons are concentrated close to the inner edge of the
accretion flow, and the reduced physical extent of the relevant 
electron population reduces the time delay between the infrared and X-ray 
peaks and eliminates the long tail in the original X-ray light curve.

\section{Discussion \& Conclusions}

The calculations presented above do not include relativistic effects
such as lensing, time delays, gravitational redshift, and relativistic
beaming of upscattered photons by the bulk motion of the material.
All of these effects are important close to the inner edge of the
accretion flow and will make a concomitant contribution to the 
shaping of the lightcurve.   In addition, the relative geometry of 
teh infrared flare source location, the black hole and the observers' line 
of sight are essentially free parameters.

Nevertheless the point remains: upscattering of infrared flare photons
by thermal electrons in the accretion flow enveloping Sgr A* will
produce significant X-rays.  The counterargument by Dodds-Eden et al.\
(2009, and elsewhere in this volume) that inverse comptons cattering
is produced only weak X-ray fluxes breaks down because the radio to
sub-millimeter emission from the accretion flow is optically thick, so
there are more electrons available to upscatter near-infrared photons to X-ray
energies than indicated simply based on the millimeter flux from Sgr
A*.  Typical accretion flow models suggest X-ray flares with peak
luminosities of $\sim 10^{33}$--$10^{34}$\,erg\,s$^{-1}$ per mJy of
peak flux in the near-infrared, as previously estimated by Yusef-Zadeh
et al.\ 2009).

Independent of whether this process is, or is not, responsible for the
observed X-ray flares, models for the accretion flow must not
overproduce X-rays, nor predict observable time delays between X-ray
and near-infrared flaring in excess of several minutes, nor overproduce extended
decay tails.  This suggests that the infrared flares occurs close to the
inner edge of Sgr A* and that the electron temperature with increasing
radius falls reasonably rapidly.

Future high-resolution imaging of Sgr A* in the near-infrared with, for example,
VLTI/GRAVITY (see the papers by Paumard et al.\ and Vincent et al.\
elsewhere in these proceedings) has the potential to strongly
constrain the location of the near-infrared flare source.  In conjunction with
simultaneous monitoring for X-ray flaring this would sharply constrain
models for the accretion flow.

\acknowledgements
Thanks to BP Pandey, Charles Gammie, and Farhad Yusef-Zadeh for 
useful discussions. This work was supported by the Australian 
Research Council through Discovery Project Grant DP0986386.

{}

\end{document}